# Design of a Scalable 4G Portable Network Using Low Cost SDR And Raspberry Pi

NUPUR CHOUDHURY, CHINMOY KALITA, KANDARPA KUMAR SARMA
Department of Electronics and Communication Engineering, Gauhati University, Guwahati, INDIA

Abstract—Of late, Software Defined Radio (SDR) approach has become an effective means to design high data rate wireless systems for a range of applications. There are methods with which low cost SDR based 4th generation (4G) or long term evolution (LTE) systems can be designed. Using low cost Raspberry Pi systems, the SDR aided 4G systems can be designed for high data rate communication. The work is related to the design of a 4G wireless system using low cost SDR solutions and integrated to a programmable controller based on a Raspberry Pi. Experimental results show that the system is effective in a range of conditions.

## 1. Introduction

The 4G remote uses Orthogonal Frequency Division Multiplexing (OFDM), Ultra Wide Radio Band (UWB), and Millimeter wireless and brilliant reception apparatus. Information rate of 20mbps is utilized. Portable speed will be up to 200km/hr. Frequency band is 2 - 8 GHz. It gives the capacity for overall wandering to get to cell anyplace. This thought was past the creative ability of common versatile client promising "interface whenever, in any case, anyplace". This universal system access will be accomplished via flawlessly coordinating the accessible and new systems utilizing a core IP based network layer.

Over the last decade, a new technology called Software Defined Radio (SDR) has emerged to tackle the problem of inter-operability with legacy and incompatible hardware radios. SDR solves the problem of inter-operability by implementing a large part of radio functionality in software rather than hardware [5]. SDR can flawlessly communicate with numerous contradictory radios or go about as a bridge between them. An SDR is a radio communication system wherein the hardware functions like filters, amplifiers, modulators etc. are implemented in software by using an FPGA, commodity computers or embedded devices [6]. By doing so, hardware complexity can be widely reduced, and in addition to software part there is a RF front end processing this software section. This design gives us flexibility, such that one can transmit and receive variety of signal waveforms based on the applications using the same hardware. SDR system can be easily operated in any environment because of its adaptability and re-programmable nature. The system can receive new frequency channels easily through software modifications. New modulation schemes can be easily adopted in SDR since it collects complex signals in digital form with various sampling frequency. Software Defined Radio (SDR) alludes to the innovation where in programming modules running on a conventional equipment stage comprising of DSPs and general-purpose microprocessors are utilized to actualize radio capacities, such as, generation of transmitted signal at transmitter (modulation) and tuning/identification of received radio signal at receiver (demodulation). Hence to build a 4G network with the ease of SDR will be a very effective process in terms of cost, portability, size etc. The model designed to build a scalable 4G network is of low cost, low power consumption and can be deployed easily during the time of needs or emergencies.

## 2. Background of SDR

The wireless communications are getting developed day by day every electronic devices are getting into the world of wireless communication In various forms like Wifi, Bluetooth and many others. They have their own limitations and advantages in different fields. Radios configured with hardware are very hard to modify. Consequently, this static nature gives hardware radios several limitations. These hardware setups use lot of space and different technologies. This becomes expensive for using various hardware protocols. Cellular phone technology is a great example of this system but due to some limitations like it needs different systems for different conversations the cost gets increases. As the hardware-based radios are not easily modified these limitations cause various problems so the engineers came out with an idea of developing radios which are based on software which are called the SDRs (Software Defined Radio). The main feature of the newly modified SDRs is to make the work easy by having various systems in converted to one system. In the past years microprocessors helped the engineers to send and receive information from one place to another. These can be used as digital way or analog way. Digital systems have lot of benefits and advantages still it couldn't replace the analog systems as they are simple to use. Digital systems may not replace the analog systems but it is seen that the use of digital systems in getting increased day by day. In view of the confinements innate in static equipment radio frameworks, an alternate sort of radio framework has been created inside the previous couple of years. To take care of the equipment issue, engineers chose to actualize parts of the radio utilizing programming instead of equipment. Utilizing programming instead of equipment to execute a few phases of a radio framework empowers a radio to be all the more effectively arranged, adjusted, and created for different frameworks. This new type of radio usage came to be known as Software Defined Radio (SDR) or programming radio. The objective of SDRs is to execute completely utilitarian radios in a single framework that already required different frameworks. The relocation from equipment characterized radios to programming characterized radios compares with the move from analog radio frameworks to digital radio framework. Before, the ascent of microchips empowered correspondences architects to build up another approach to transmit data starting with one place then onto the next. These correspondence digital frameworks gave some key advantages over the analog correspondence frameworks. Notwithstanding the points of interest, in any case, both analog and digital correspondence frameworks still exist today. Since both analog and digital system are useful in the communication system but for digital communication system it is not easy to remove completely the analog system. Parallelly, it will be seen that the use of the software radios may allow them to replace all the hardware devices in many applications over the coming years; be that as it may, the effortlessness and constancy of equipment radios will guarantee these radios keep on existing, too. Similarly as radio frameworks once experienced a period of changing over from simple to advanced, radio frameworks today and later on are progressively getting to be programming characterized as opposed to equipment characterized.

## 3. Literature Review

In the defense sector SDR have been known since the late 1970s in both the U.S. and Europe [1]. The first military application that uses SDR was Speak Easy program that was initiated by the U.S. military. The program aimed to use programmable processing o emulate multiple existing radios, and to easily incorporate new coding and modulation standards so that military communication can keep pace with advances with coding and modulation technique. Another program by the U.S. military in the Joint Tactical radio system (JTRS) which produces a family of interoperable, modular, software defined radio that operate as nodes in a network that provides secure wireless communication and networking for mobile and fixed forces, consisting of U.S. Allies, joint and coalition partner, and in time, disaster response personnel .Over the last two decades there have been many application and research publications on SDR [2-11].

There are several companies that develop the hardware or the software products to be used for different applications such that Flex radio system, SunSDR, BladeRF. Open source hardware is also available for software defined radio.

Many current SDR platforms are based on programmable hardware such as field programmable gate arrays (FPGAs) or embedded digital signal processors (DSPs).

Such hardware platforms can meet the processing and timing requirements of modem high –speed wireless protocols, but programming FPGAs and specialized DSPs can be a difficult tasks. Developers have to learn how to program for eachcomponents of embedded architecture, often without the support of a rich development environment of programming and debugging tools. Additionally, such specialized hardware platforms can also be expensive, e.g., at least several times the cost of an SDR platform based on a general- purpose processor (GPP) architecture, such as general purpose personal computers (PC). A GPP based SDR offers maximum flexibility and the easiest development [7]. GPPs are the best suited for SDRs used for handheld applications, because size and power consumptions are significantly less, and the ability to quickly implement a range of algorithms and waveforms is easier. SDR platforms that use general-purpose PCs enable developers to use a familiar architecture and environment having numerous sophisticated programming and debugging tools available.

There has been few patents around the developments of SDR, for instance [13] discloses a processing platform in compact size capable of inter alia transmitting or receiving radio signals in a large frequency bands. The device disclosed uses external logical boards, digital to analog converters, up-converter, and mixers apart from a central processing platform for the transmission of radio signals in the large frequency band. In another patent [14], represent a single module having two transceivers. A core processing element inside each of the two software defined radio transceiver has total dissipated power (TDP) rating of less than 2 watts, thereby enabling operation without overheating. The core processor in some of the embodiments is an OMAP processor. Each of the SDR transceiver can be operated or transitioned between different types of waveforms.

## 4. Overall Hardware Architecture

Figure:1 demonstrates the overall block diagram of the system. The system incorporates a Raspberry Pi board, RTL-SDR, an antenna, a speaker, a monitor display and a receiver. Radio signal are received utilizing an antenna associated with RTL-SDR. RTL-SDR is a ware USB control SDR receiver only. Analog signal which is received by antenna is sent to analog to digital converter (ADC) segment of RTL-SDR, where analog signal is digitized at a sampling rate which is set by software. Digitized radio signal is then down converted to frequency which is again controllable by the software.

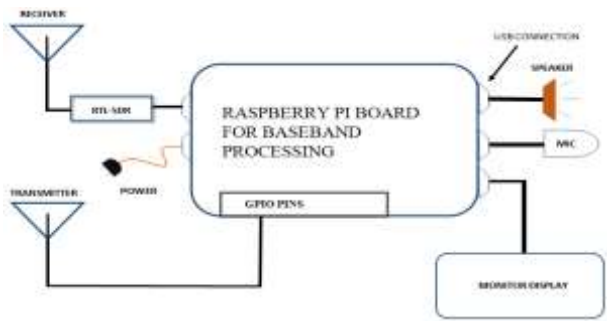

Figure:1. Hardware Architecture

Base band signal is then sent to Raspberry pi for extraction of information from the signal. The data extracted can either be channeled to the speaker associated with the system or steered to the memory for capacity. With the end goal to transmit the signal, general purpose input output (GPIO) pin accessible on Raspberry pi is utilized, in this way expelling the requirement for committed radio frequency (RF) equipment.

In our overall framework it comprises of three noteworthy equipment segments, they are receiver, baseband processor and transmitter.

### 4.1 Receiver

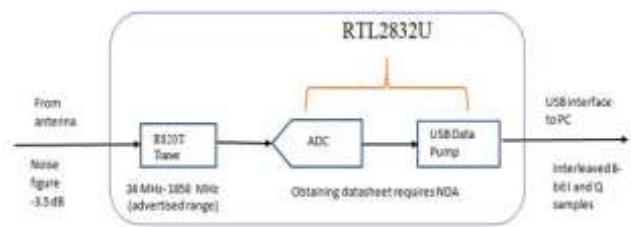

Figure: 2. Architecture of RTL-SDR

For execution of receiver front end, a low-cost DVB-T dongle based on the Realtek RTL2832U is used [8]. This chip permits exchanging the raw I/Q samples to the associated device. The RTL-SDR dongle essentially contains two chips to be specific; the Raphael Micro R820T radio tuner and the Realtek RTL2832U (contains a 8 bit ADC and USB data pump). The essential chip setup is portrayed in the block diagram of Figure: 2. The tuner chip fills in as the RF front end for the RTL 2832U receiver. The receiving wire is associated by a small scale cajole. Low Noise Amplifier (LNA), which pursues the cajole connector, gives a noise figure (NF) of around 3.5 dB. The tuning scope of the R820T is 24MHz to 1850 MHz and utilizes a 3.75 MHz intermediate frequency. Reception apparatus antenna input is 75 ohm impedance. The dynamic range is around 45dB. The most elevated example rate is 2.4 MS/s. Down changing over the received RF to an intermediate frequency (IF) is done by frequency synthesizer inside the R820T, creates a local oscillator (LO) signal. 1 Hz of tuning goals and gain can be arranged by programming control. Receiver digital signal processing is finished by this module, which incorporates down converting and additional filtering of the IF signal conveyed by the R820T.

The ADC produces 8 bit real (in phase signified by I) and imaginary (quadrature signified by Q) interleaved samples esteems, in an unsigned format. The RTL SDR dongle utilizes a phase locked loop based synthesizer to deliver the local oscillator required by the quadrature mixer. The actual output is interleaved; so one byte I, then one byte Q with no header or metadata (timestamps). The samples themselves are unsigned and are in the range from 0 to 255. The RTL2832U likewise contains a USB interfaced that sends samples to Raspberry pi.

### 4.2 Baseband Processor

For the baseband processing of the signal, we utilized a low power credit card size single board PC Raspberry pi demonstrate B (Figure: 4) [9-15]. The CPU on the load up is an ARM processor with 700 MHz clock speed. CPU execution can be contrasted with a Pentium II 300 MHz processor and the GPU execution is like the original Xbox. It has an assortment of interfacing peripherals, counting USB port, HDMI port, 512 MB RAM, SD card stockpiling and interestingly GPIO ports for extension. Monitor, keyboard and mouse can be associated with Raspberry pi through HDMI and USB connectors and it very well may be utilized like a Desktop PC. It bolsters various working framework including a Debian based Linux Distro, Raspbian, which is utilized in our structure. Raspberry Pi can be associated with a local area network through Ethernet link or USB WIFI connector, and after that it very well may be gotten to through SSH remote login. ARM processor are ended up being essentially better regarding power, cost and size then other competing gadgets, thus shape core processor of most versatile and implanted frameworks.

Raspberry Pi is associated with RTL-SDR utilizing USB connection. I/Q samples received from RTL-SDR are processed on Raspberry pi utilizing the demodulation software program written in C programming dialect. This program makes utilization of open source driver accessible for RTL SDR which can be utilized to control the frequency, bandwidth and gain of received signals. Different sort of modulated signals like Amplitude Modulation (AM), single side band, Frequency Modulation (FM), Wide band FM, narrow band FM are executed on Raspberry pi which can be utilized to received different signals accessible broadcasting live. The demodulated signal is then funneled to USB amplifier associated with Raspberry Pi.

### 4.3 Transmitter

We present a novel design to implement a transmitter by utilizing GPIO pins on Raspberry pi board, which spares us We present a novel design to implement a transmitter by utilizing GPIO pins on Raspberry pi board, which spares us from having additional RF hardware equipment.

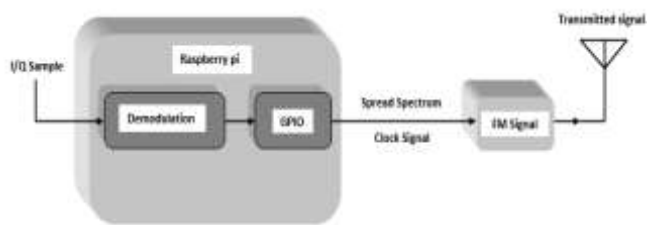

Figure: 3. Function of GPIO pin

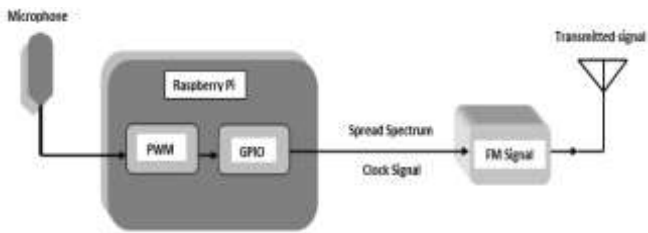

Figure: 4. Function of PWM Hardware

Figure: 3 shows the hardware on the Raspberry pi that is really intended to create spread spectrum clock signal on GPIO pins is used to produce FM signal. Raspberry pi is set up to output a waveform corresponding to the central frequency of FM signal. Then Figure: 4 shows the PWM hardware on Raspberry pi is used to change the frequency of output signal according to the amplitude of the audio received from the microphone. Accordingly the output is a waveform whose frequency is proportional to the amplitude of the input audio. In this way, a FM transmitter is implemented by simply utilizing the pulse width modulation (PWM) hardware and GPIO pins of Raspberry Pi. The frequency of transmission of the FM transmitter is software configured and hence provides the flexibility to the tuned to any crisis/aid services receivers frequency in real time.

## 5. Operation

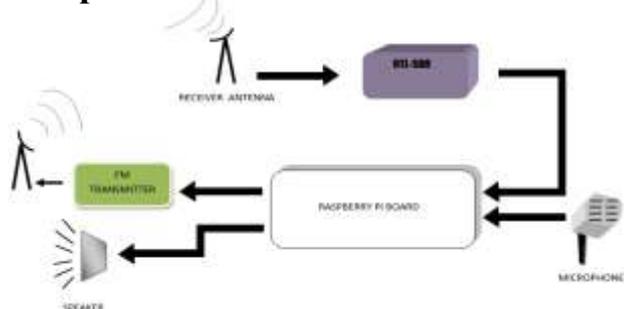

Figure: 5. Operation of our model

A method for software defined communication utilizing transceiver module, containing designing at least one processor to receive a transmission of a radio signal in a substantial frequency band or gathering of the radio signal from the large frequency band was clarified in past segment. In this segment, we describe the activity of the equivalent. We initially instate an hardware module design to receive or transmit the radio signal, wherein a VCO module transmit the radio signal as a defined frequency ; and transmitting the radio signal over an antenna utilizing a transmitting module. The software module initially receives an input instruction for type of modulation, a second input instruction for frequency, third input signal through audio input connected to one of said USB port. The receive radio signal is sent to at least one audio output associated with the Raspberry pi board through USB port, where in the receive radio signal can be stored in said storage unit, where in the hardware module controls the radio signals gathering associated through at least one USB port. For transmission, the software module changes the frequency of the input signal in accordance to amplitude of said audio input using pulse width modulation and transmitting the modulated radio signal through the said transmitter.

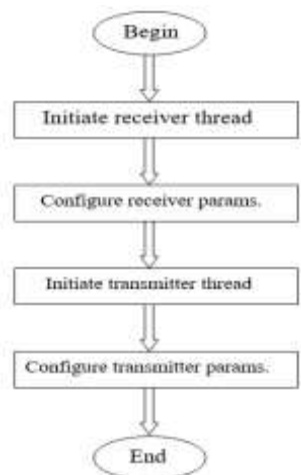

Figure: 6. Flowchart of the operation of the device

Figure: 5 demonstrates flowchart of the operation of the device. Firstly, the user starts the recipient strings by running the receiver program where the frequency of operation and type of modulation scheme are provided as parameters. Additionally, user starts the transmitter program and gives frequency of transmission as a contention to the program. User can change the frequency of operation and modulation for both reception and transmission, in this manner giving adaptability to be tuned to different frequencies and protocols in real time. User can transmit his voice/audio signal as in traditional radios or transmit recorded signals. So also, it can store the receive signals on a memory card or route to an external device.

This device is capable of working in broad range of frequencies. In this way, it is conceivable to utilize it to communicate with different set of devices by changing its configuration which can be controlled using software. Figure: 6 depicts one of the conceivable methods for utilizing these transceivers to communicate with different devices operating in multiple frequency bands. This transceiver can communicate with various arrangements of crisis/help offices radios by changing its configuration, which can be control using software. Figure: 6 depicts one of the conceivable methods for utilizing this transceiver to communicate with different devices operating in multiple frequency bands.

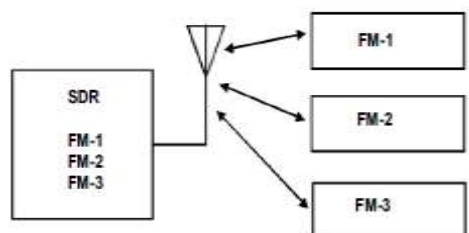

Figure:7. Communication with devices operating in different frequency bands.

In this Figure: 7, FM - 1 signifies the first device working on frequency f1, similarly FM-2 denotes the first device working on frequency f2 and FM-3 denotes the first device working on frequency f3.

# 6. Performance Analysis

Utilizing Raspberry pi in the frontend has helped in decreasing the power requirement as observed in the larger gadgets. Power consumed by this processor is under 3 watts. Utilization of RTL-SDR gives a minimal effort front-end arrangement. The arrangement is convenient and can be immediately designed to fulfill immediate requirements mainly during emergency situations. Further, the proposed approach is a low cost one.

A comparison of the results derived using various combinations available in the RTL-SDR is shown in the Table 1.

NFM: Narrowband Frequency Modulation.
WFM: Wideband Frequency Modulation.
AM: Amplitude Modulation
DSB: Double Side Band.
Order: 1000, Gain: 19dB,
Sampling rate: 2.4MSPSs
Sampling Mode: Quadrature Sampling.

| Baseband Frequency (Hz) | Stepsize (kHz) | Filter | Modulation Scheme | Quality Strong | Medium | weak |
|---|---|---|---|---|---|---|
| 8000 | 12.5 | Blackman-Harris 4 | NFM | SNR=24.6dB | | |
| 8000 | 12.5 | Blackman-Harris 4 | WFM | | SNR=14.6dB | |
| 8000 | 12.5 | Blackman-Harris 4 | AM | | | SNR=4.0dB |
| 8000 | 12.5 | Blackman-Harris 4 | DSB | | | SNR=3.6dB |
| 10000 | 5 | Blackman-Harris 7 | NFM | | SNR=15.7dB | |
| 10000 | 5 | Blackman-Harris 7 | WFM | | SNR=12.6dB | |
| 10000 | 5 | Blackman-Harris 7 | AM | SNR=27.6dB | | |
| 10000 | 5 | Blackman-Harris 7 | DSB | | | SNR=4.2dB |
| 116240 | 10 | Blackman | NFM | | SNR=19.6dB | |
| 116240 | 10 | Blackman | WFM | SNR=34.6dB | | |
| 116240 | 10 | Blackman | AM | | SNR=15.6dB | |
| 116240 | 10 | Blackman | DSB | | | SNR=7.6dB |
| 6000 | 0.01 | Blackman-Harris 4 | NFM | | | SNR=3.7dB |
| 6000 | 0.01 | Blackman-Harris 4 | WFM | | | SNR=1.9dB |
| 6000 | 0.01 | Blackman-Harris 4 | AM | | SNR=15.2dB | |
| 6000 | 0.01 | Blackman-Harris 4 | DSB | SNR=28.6dB | | |

Table 1: Quality analysis of received signal.

The experimental results show that the RTL-SDR platforms enables the receive of signals with better SNR and quality with the flexibilities of using multiple combinations of baseband frequency, modulation scheme, filter types and step size with no modification to hardware.

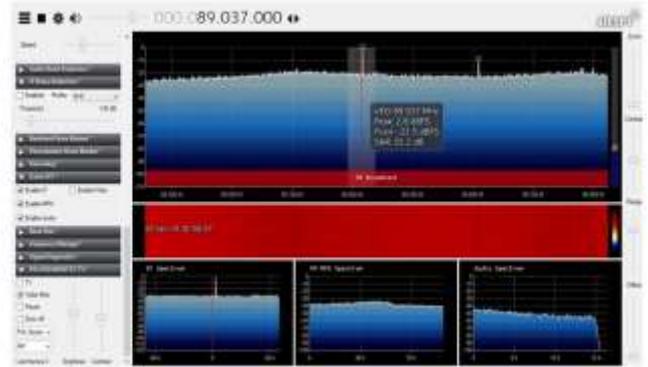

Figure: 8. Received FM signal and its different spectrums in SDR sharp.

The received FM signal shown in Figure: 8. Where we can observe SNR value of signal and various spectrum analysis of receive signal.

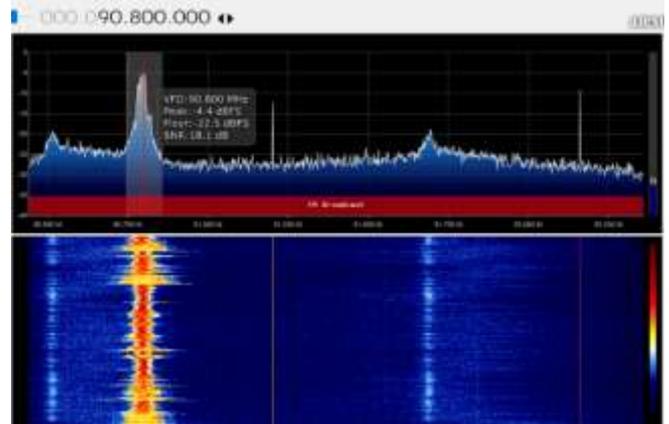

Figure: 9. Received FM broadcast signal by RTL-SDR in SDR-SHARP platform.

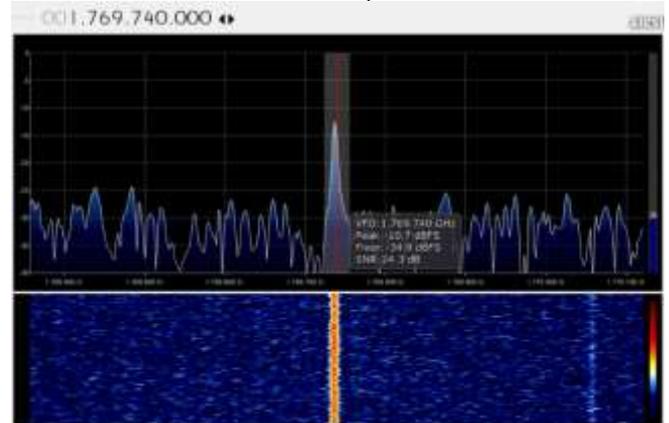

Figure: 10. Received GSM signal by RTL-SDR in SDR-SHARP platform

Another FM broadcasting signal at 90.8 MHz and Global System for Mobile Communication (GSM) signals by RTL-SDR are received and its spectrum and spectrogram are shown in Figure: 9 and Figure: 10 respectively. The red portion in this spectrogram shows the strength of the received signals. The signals which are available and are in the range of antenna can be received with our RTL-SDR without any further changes to the software as well as hardware.

# 7. Conclusion

In this paper, we have discussed the design of a low cost, efficient and portable framework for deploying a 4G/LTE network using RTL-SDR interfaced with a controller built on a Raspberry pi. The proposed approach is found to be reliable and is found to be suitable for transmission of audio and video signals. The quality of reception is also satisfactory. In an extended form, the present approach may be used to construct a LTE network in an emergency situation.